\documentclass[aps,prd,onecolumn,groupedaddress,showpacs,nofootinbib,amssymb]{revtex4}
\usepackage[dvips]{graphicx}
\usepackage{amssymb}
\usepackage{amsmath}
\usepackage{hyperref}
\usepackage{graphicx,color}
\usepackage{amsfonts}
\usepackage{bm}
\usepackage{cancel}
\usepackage{comment}

\newcommand\be{\begin{equation}}
\newcommand\ee{\end{equation}}

\allowdisplaybreaks[4]

\begin{document}

\tolerance=5000

\title{Static Neutron Stars Perspective of Quadratic and Induced Inflationary Attractor Scalar-tensor Theories}

\author{V.K. Oikonomou,$^{1,2}$}
\email{voikonomou@auth.gr,voiko@physics.auth.gr,v.k.oikonomou1979@gmail.com}
\affiliation{$^{1)}$ Department of Physics, Aristotle University
of Thessaloniki, Thessaloniki 54124, Greece}
\affiliation{$^{2)}$L.N. Gumilyov Eurasian National University -
Astana, 010008, Kazakhstan}

\tolerance=5000

\begin{abstract}
This study focuses on the static neutron star perspective for two
types of cosmological inflationary attractor theories, namely the
induced inflationary attractors and the quadratic inflationary
attractors. The two cosmological models can be discriminated
cosmologically, since one of the two does not provide a viable
inflationary phenomenology, thus in this paper we investigate the
predictions of these theories for static neutron stars, mainly
focusing on the mass and radii of neutron stars. We aim to
demonstrate that although the models have different inflationary
phenomenology, the neutron star phenomenology predictions of the
two models are quite similar. We solve numerically the
Tolman-Oppenheimer-Volkoff equations in the Einstein frame using a
powerful double shooting numerical technique, and after deriving
the mass-radius graphs for three types of polytropic equations of
state, we derive the Jordan frame mass and radii. With regard the
equations of state we use polytropic equation of state with the
small density part being either the WFF1, the APR or the
intermediate stiffness equation of state SLy. The results of our
models will be confronted with quite stringent recently developed
constraints on the radius of neutron stars with specific mass. As
we show, the only equation of state which provides results
compatible with the constraints is the SLy, for both the quadratic
and induced inflation attractors. Thus nowadays, scalar-tensor
descriptions of neutron stars are quite scrutinized due to the
growing number of constraining observations, which eventually may
also constrain theories of inflation.
\end{abstract}

\pacs{04.50.Kd, 95.36.+x, 98.80.-k, 98.80.Cq,11.25.-w}

\maketitle

\section*{Introduction}

The current and future status of theoretical cosmology and
astrophysics is updated day by day, since astrophysical and
cosmological observations are offered in high frequency. For
astrophysics, the starting point of profound importance was the
kilonova merging event known as GW170817
\cite{TheLIGOScientific:2017qsa,Abbott:2020khf} in which a direct
observation of gravitational waves was first recorded. After that
event, neutron stars (for reviews and textbooks see
\cite{Haensel:2007yy,Friedman:2013xza,Baym:2017whm,Lattimer:2004pg,Olmo:2019flu})
and other astrophysical compact objects enjoy an elevated
scientific role. Specifically, neutron stars are at the crossroads
of many scientific eras, like nuclear theory
\cite{Lattimer:2012nd,Steiner:2011ft,Horowitz:2005zb,Watanabe:2000rj,Shen:1998gq,Xu:2009vi,Hebeler:2013nza,Mendoza-Temis:2014mja,Ho:2014pta,Kanakis-Pegios:2020kzp,Tsaloukidis:2022rus},
high energy physics
\cite{Buschmann:2019pfp,Safdi:2018oeu,Hook:2018iia,Edwards:2020afl,Nurmi:2021xds},
modified gravity
\cite{Astashenok:2020qds,Astashenok:2021peo,Capozziello:2015yza,Astashenok:2014nua,Astashenok:2014pua,Astashenok:2013vza,Arapoglu:2010rz,Panotopoulos:2021sbf,Lobato:2020fxt,Numajiri:2021nsc}
and astrophysics
\cite{Altiparmak:2022bke,Bauswein:2020kor,Vretinaris:2019spn,Bauswein:2020aag,Bauswein:2017vtn,Most:2018hfd,Rezzolla:2017aly,Nathanail:2021tay,Koppel:2019pys,Raaijmakers:2021uju,Most:2020exl,Ecker:2022dlg,Jiang:2022tps}.
The role of modified gravity in neutron star physics is still
questioned and under investigation. To date, the need of modified
gravity descriptions for the large scale structure of our Universe
seems rather compelling, since dark energy cannot be described in
a consistent way by the use of General Relativity and scalar
fields solely. To date, many works have been devoted on the
physics of neutron stars in the context of modified gravity
\cite{Astashenok:2014nua,Astashenok:2014pua} and its scalar-tensor
counterpart theories
\cite{Pani:2014jra,Staykov:2014mwa,Horbatsch:2015bua,Silva:2014fca,Doneva:2013qva,Xu:2020vbs,Salgado:1998sg,Shibata:2013pra,Arapoglu:2019mun,Ramazanoglu:2016kul,AltahaMotahar:2019ekm,Chew:2019lsa,Blazquez-Salcedo:2020ibb,Motahar:2017blm,Odintsov:2021qbq,Odintsov:2021nqa,Oikonomou:2021iid,Pretel:2022rwx,Pretel:2022plg,Cuzinatto:2016ehv}.
In scalar-tensor theories, there exists a subclass of inflationary
phenomenological models, which have the spectacular property of
being inflationary attractors in the Einstein frame
\cite{alpha0,alpha1,alpha2,alpha3,alpha4,alpha5,alpha6,alpha7,alpha7a,alpha8,alpha9,alpha10,alpha11,alpha12,alpha13,alpha14,alpha15,alpha16,alpha17,alpha18,alpha19,alpha20,alpha21,alpha22,alpha23,alpha24,alpha25,alpha26,alpha27,alpha28,alpha29,alpha30,alpha31,alpha32,alpha33,alpha34,alpha35,alpha36,alpha37}.
Specifically, distinct theories in the Jordan frame where the
scalar field is non-minimally coupled to gravity, lead to Einstein
frame counterpart theories which have the same inflationary
observational indices, namely the same expressions for the
spectral index of the primordial scalar curvature perturbations
and the same tensor-to-scalar ratio. Most of the distinct
attractor models are compatible with the Planck data
\cite{Akrami:2018odb}, thus most of these are cosmologically
viable theories. In this work we shall be interested in two types
of distinct cosmological attractors, the quadratic attractors and
the induced inflation attractors, see \cite{alpha0} for a
description of both the models. These two models correspond to
different limiting cases of the coupling of the scalar field to
gravity in the Jordan frame. We are interested in examining the
implications of such attractor potentials on static neutron stars.
After describing in brief the essential cosmological features of
the two distinct attractors in the Jordan and Einstein frames, we
shall derive the Einstein frame Tolman-Oppenheimer-Volkoff, which
we shall solve numerically using a powerful double shooting method
in order to obtain optimal initial conditions at the center of the
star. The numerical method is a python 3 LSODA integrator
technique which relies on double shooting for the central values
of the scalar field and the metric function \cite{niksterg}. The
integration shall be performed for three distinct piecewise
polytropic \cite{Read:2008iy,Read:2009yp} equations of state, the
low density part of which shall be the Wiringa-Fiks-Fabrocini
(WFF1 hereafter) \cite{Wiringa:1988tp}, the Skyrme-Lyon (SLy
hereafter) \cite{Douchin:2001sv}, or the
Akmal-Pandharipande-Ravenhall (APR hereafter) equation of state
\cite{Akmal:1998cf}. The numerical analysis will yield the
Einstein frame radii and Arnowitt-Deser-Misner (ADM) masses
\cite{Arnowitt:1960zzc} of neutron stars, so by using appropriate
formulas we derive the Jordan frame masses and radii. For all the
cases studied we will confront the outcomes with stringent
observations and constraints of specific mass neutron stars radii.
We shall consider three types of constraints, developed in Refs.
\cite{Altiparmak:2022bke,Raaijmakers:2021uju,Bauswein:2017vtn}. As
we show, the constraints eliminate completely the APR and WFF1
descriptions of neutron stars, and only the SLy equation of state
satisfies all the constraints. This work is aligned with the new
physics era discipline, where constraints on astrophysical objects
may affect to some extent cosmological theories. In our case, both
the inflationary attractors can describe static neutron stars
consistently, however only when the matter is described by an
intermediate stiffness equation of state. This is contrary to
previous works, where the inflationary attractors could provide a
viable description of neutron stars for all the equations of state
we discuss in this paper, see for example
\cite{Odintsov:2021qbq,Odintsov:2021nqa,Oikonomou:2021iid}.

\section{Induced and Quadratic Inflationary Attractors and their Inflationary Phenomenology}

The quadratic and induced inflationary attractors belong to a much
more general class of cosmological inflationary attractors
\cite{alpha0,alpha1,alpha2,alpha3,alpha4,alpha5,alpha6,alpha7,alpha7a,alpha8,alpha9,alpha10,alpha11,alpha12,alpha13,alpha14,alpha15,alpha16,alpha17,alpha18,alpha19,alpha20,alpha21,alpha22,alpha23,alpha24,alpha25,alpha26,alpha27,alpha28,alpha29,alpha30,alpha31,alpha32,alpha33,alpha34,alpha35,alpha36,alpha37},
and constitute basically two different non-minimal coupling limits
of the same Jordan frame non-minimally coupled scalar-tensor
theory. Specifically, the gravitational action of the common
Jordan frame non-minimal coupled theory is,
\begin{equation}\label{c1}
\mathcal{S}_J=\int
d^4x\sqrt{-g}\Big{[}\Omega(\phi)R-\frac{1}{2}g^{\mu
\nu}\partial_{\mu}\phi\partial_{\nu}\phi-U(\phi)\Big{]}+S_m(g_{\mu
\nu},\psi_m)\, ,
\end{equation}
where $\psi_m$ denotes the perfect matter fluids in the Jordan
frame, which have pressure $P$ and energy density $\epsilon$, and
$g$ is the determinant of the Jordan frame metric $g_{\mu \nu}$.
Note that for the inflationary models which shall be considered in
this work, the kinetic term of the scalar field has a canonical
minimal coupling and appears as $-\frac{\omega(\phi)}{2}g^{\mu
\nu}\partial_{\mu}\phi\partial_{\nu}\phi$ and not as
$-\frac{1}{2}g^{\mu \nu}\partial_{\mu}\phi\partial_{\nu}\phi$.
Also in the case at hand the non-minimal coupling function in the
Jordan frame $\Omega(\phi)$ and the scalar potential $U(\phi)$ has
the following form \cite{alpha0},
\begin{equation}\label{nnminimalandpotential}
\Omega(\phi)=\left(1+\xi f(\phi)
\right),\,\,\,U(\phi)=M_p^4\lambda \left(\Omega(\phi)-1
\right)^2\, ,
\end{equation}
where $M_p=\frac{1}{\sqrt{8\pi G}}$ is the reduced Planck mass and
$G$ is Newton's gravitational constant. The parameters $\lambda$
and $\xi$ are dimensionless in natural units, and also the same
applies for the non-minimal coupling function $f (\phi)$. Also we
shall refer to the parameter $\xi$ as the non-minimal coupling
hereafter. The induced and quadratic inflationary attractors
correspond to different limits of the non-minimal coupling $\xi$,
namely the strong coupling limit $\xi\gg 1$ and the weak coupling
limit $\xi\ll 1$ respectively. Let us consider the essential
inflationary features of the induced inflationary attractors, so
the strong coupling limit of the initial theory. In this case, the
non-minimal coupling function $\Omega (\phi)$ and the potential
take the form,
\begin{equation}\label{inducedpotjordan}
\Omega (\phi)=\xi f(\phi),\,\,\,U(\phi)=\lambda M_p^4\xi^2
f(\phi)^2\, .
\end{equation}
We can obtain the Einstein frame theory which has the inflationary
attractor property by performing an appropriate conformal
transformation \cite{Kaiser:1994vs,valerio},
\begin{equation}\label{c4}
\tilde{g}_{\mu \nu}=\frac{M_p^2}{2}\Omega^2g_{\mu \nu}\, ,
\end{equation}
with the Einstein frame metric being $\tilde{g}_{\mu \nu}$, and
the ``tilde'' will indicate the Einstein frame quantities
hereafter. The minimally coupled Einstein frame canonical scalar
theory action reads,
\begin{equation}\label{c12}
\mathcal{S}_E=\int
d^4x\sqrt{-\tilde{g}}\Big{[}\frac{M_p^2}{2}\tilde{R}-\frac{1}{2}
\tilde{g}^{\mu \nu }\tilde{\partial}_{\mu}\varphi
\tilde{\partial}_{\nu}\varphi-V(\varphi)\Big{]}+S_m(\Omega^{-2}\frac{2}{M_p^2}\tilde{g}_{\mu
\nu},\psi_m)\, ,
\end{equation}
and the Einstein frame scalar potential of the canonical scalar
field $\varphi$, namely $V(\varphi)$ takes the following form,
\begin{equation}\label{c13}
V(\varphi)=V_s\left(1-e^{-\sqrt{\frac{2}{3}}\frac{\varphi}{M_p}}
\right)^2\, ,
\end{equation}
where $V_s=\frac{M_p^4\lambda}{\xi^2}$, and also in terms of the
canonical scalar field, the non-minimal coupling function $\Omega
(\varphi)$ takes the form,
\begin{equation}\label{nonmimimalinduced}
\Omega (\varphi)=e^{\sqrt{\frac{2}{3}}\frac{\varphi}{M_p}}\, .
\end{equation}
Also note that the matter fluids in the Einstein frame are not
perfect fluids, since their energy-momentum tensor satisfies the
following continuity equation,
\begin{equation}\label{c24}
\tilde{\partial}^{\mu}\tilde{T}_{\mu \nu}=-\frac{d}{d\varphi}[\ln
\Omega]\tilde{T}\tilde{\partial}_{\nu}\phi\, .
\end{equation}
The Einstein frame amplitude $\Delta_s^2$ of the canonical scalar
field scalar perturbations,
\begin{equation}\label{scalaramp}
\Delta_s^2=\frac{1}{24\pi^2}\frac{V(\varphi_f)}{M_p^4}\frac{1}{\epsilon(\varphi_f)}\,
,
\end{equation}
is constrained to be \cite{Akrami:2018odb},
\begin{equation}\label{scalarampconst}
\Delta_s^2=2.2\times 10^{-9}\, ,
\end{equation}
where $\varphi_f$ is the value of the scalar field at the end of
the inflationary era. Thus the parameter $V_s$ is constrained to
be,
\begin{equation}\label{tilde}
V_s\sim 9.6\times 10^{-11}\, M_p^4\, .
\end{equation}
Hence the induced inflation parameters $\lambda$ and $\xi$ can be
chosen in such a way so that $\frac{\lambda}{\xi^2}\sim 10^{-11}$,
so a large $\xi\sim 10^5$ and a $\lambda \sim 1$ can sufficiently
produce a viable inflationary era. For the induced inflationary
attractors case, the spectral index of the primordial scalar
curvature perturbations $n_s$ and the ratio of the tensor
perturbations over scalar perturbations in terms of the
$e$-foldings number $N$ take the form,
\begin{equation}\label{spectralindexsmallalpha}
n_s=1-\frac{2}{N}\, ,\,\,\,r=\frac{12}{N^2}\, .
\end{equation}
Apparently, the phenomenology of the induced inflationary scenario
is identical to the Starobinsky $R^2$ model. In fact, even the
potential (\ref{nonmimimalinduced}) is identical to the
Starobinsky model, and this justifies the terminology attractors
for the induced inflation Jordan frame theories. Note that the
whole analysis is performed without even defining the form of the
function $f(\phi)$ in the Jordan frame. An arbitrary $f(\phi)$ can
lead to the same Einstein frame theory, and this justifies the
terminology inflationary attractors. The Einstein frame theory and
the corresponding phenomenology constitutes the inflationary
attractor theory, regardless the exact functional form of the
Jordan frame function $f(\phi)$. Notice that, by conformal
invariance, the Jordan frame theory and the Einstein frame ones
should produce the same inflationary phenomenology. Hence, the
induced inflationary theory in the Einstein frame has the
following final form,
\begin{equation}\label{einsteinframeaction}
\mathcal{S}_E=\int
d^4x\sqrt{-\tilde{g}}\Big{[}\frac{M_p^2}{2}\tilde{R}-\frac{1}{2}\tilde{g}^{\mu
\nu } \tilde{\partial}_{\mu}\varphi
\tilde{\partial}_{\nu}\varphi-V_s\left(1-e^{-\sqrt{\frac{2}{3}}\frac{\varphi}{M_p}}\right)^{2}\Big{]}\,
,
\end{equation}
which is identical with the Starobinsky model in the Einstein
frame.

At this point let us discuss an important issue. The inflationary
attractor models are basically non-minimally coupled scalar field
theories in the Jordan frame which yield in most cases an
identical phenomenology in the Einstein frame. For example in the
present case, the induced inflation attractors in the Einstein
frame yield an inflationary phenomenology which is identical to
the Starobinsky model in the Einstein frame. Thus these two
attractors are indistinguishable when the cosmic microwave
background inflationary modes are probed. These models provide a
viable phenomenology and always lead to the same spectral index
and tensor-to-scalar ratio which do not depend on the specific
free parameters of the models, see for example Eq.
(\ref{spectralindexsmallalpha}). However, the parameters of each
model, like the amplitude of the scalar perturbations are
constrained by the Planck data, see Eq. (\ref{scalarampconst}),
thus the coefficient of the potential \emph{is not } a free
parameter in the theory and should not be treated as being such.
In most neutron star works  in the context of non-minimally
coupled scala-tensor theories, the coefficient of the potential is
treated as a free parameter, and this is not correct. One of the
major contributions of this work is the fact that we do not
actually consider the coefficient of the potential as a free
parameter, but on the contrary we take its values to be compatible
with the cosmological constraints, see for example Eq.
(\ref{tilde}). Thus with our work we actually see the effects of a
viable inflationary theory directly on neutron stars, and we
examine what phenomenology we would get. The ultimate goal is to
see cosmological theories effects on the neutron star level. The
contrary however, is not possible, that is to constrain
cosmological theories via neutron stars. This is because we
already use a viable cosmological theory, so the free parameters
are pretty much fixed. What is possible though is to discriminate
otherwise indistinguishable cosmological theories in neutron
stars. Thus what we did in this paper is to choose the correct
inflationary non-minimally coupled theory, like in Refs.
\cite{referee1ref,Santos:2022oep} and the other inflationary
theories we mentioned in this paper, choose the parameters in the
way so that these models are viable inflationary theories, and
then apply the theory to neutron stars in order to see what kind
of neutron star predictions are obtained. This is a striking
difference between this work and already existing neutron star
physics works, in which all the parameters of the models are
chosen freely, which is incorrect from a cosmological point of
view.

Now let us turn our focus to the quadratic attractors, in which
case the parameter $\xi$ must take small values, so this is the
weak coupling limit of the Jordan frame theory describe by Eqs.
(\ref{c1}) and (\ref{nnminimalandpotential}). Thus in the case at
hand, in the weak coupling limit, the function $\Omega (\varphi)$
and the potential in the Einstein frame take the following form at
leading order \cite{alpha0},
\begin{equation}\label{quadraticfunctions}
\Omega(\varphi )=\xi\left (\xi^{-1}+\frac{g_1}{M_p}\varphi
\right),\,\,\,V(\varphi)=\lambda \frac{g_1^2}{M_p^2}\varphi^2\, ,
\end{equation}
with $g_1\ll \xi^{-1}$, and $g_1$ is the expansion parameter. The
present theory results from the weak coupling expansion in terms
of the canonical scalar field around the minimum of the theory
which is $\Omega (\phi)=1$, see \cite{alpha0} for details on the
full analysis. In the case at hand, the Einstein frame theory has
the following spectral index of primordial scalar perturbations
and tensor-to-scalar ratio,
\begin{equation}\label{spectralindexlargelalpha}
n_s=1-\frac{2}{N}\, ,\,\,\,r=\frac{8}{N}\, .
\end{equation}
Note that in this case too, the exact form of the function
$f(\phi)$ is not given, so basically many different functional
forms of $f(\phi)$ can lead to the same small coupling expansion
forms given in Eq. (\ref{quadraticfunctions}). The constraints on
the amplitude of the scalar inflationary perturbations indicate
that $\lambda g_1^2\sim 10^{-11}$ and this can easily be achieved
by choosing $\lambda \sim \mathcal{O}(1)$ and $g_1\sim
\mathcal{O}(10^{-5})$, while $\xi$ can be of the order $\xi \sim
{10^{-2}}$ in this case. As we show shortly though, the values of
$\xi$ and $g_1$ do not affect significantly the behavior of the
mass-radius diagrams. It is the functional form of the function
$\Omega (\varphi)$ that will mainly affect the neutron stars
phenomenology. Before closing, and in order to make smoother
contact with the astrophysics notation for the Einstein frame
action, we provide here a useful expression for the gravitational
action of the Einstein frame theory,
\begin{equation}\label{einsteinframeactioninflationns}
\mathcal{S}_E=\int d^4x\sqrt{-\tilde{g}}\Big{[}\frac{1}{16\pi
G}\tilde{R}-\frac{1}{2}\tilde{g}^{\mu \nu }
\tilde{\partial}_{\mu}\varphi
\tilde{\partial}_{\nu}\varphi-\frac{16\pi G V(\varphi)}{16\pi
G}\Big{]}\, ,
\end{equation}
and recall that $M_p^2=\frac{1}{8\pi G}$.

Finally let us discuss at this point an important motivational
issue regarding this work. Currently we live in the era of
precision cosmology, so many cosmological models of inflation are
put to test by either the Planck data which provide constraints on
the inflationary parameters through the cosmic microwave
background radiation. So far no evidence of B-modes in the cosmic
microwave background radiation has been found, so inflation itself
will be probed by the stage four cosmic microwave background
radiation experiments in 2027, or by the gravitational wave
experiments like LISA and the Einstein telescope in 2035. However,
neutron stars themselves may provide useful information for the
modification of gravity, if this is the correct description of
strong gravity regimes. If by studying neutron stars, it proves
that modified gravity is needed to assist the general relativistic
description, then inflationary models are in the first line of
candidate theories. The reason is that if indeed inflation ever
occurred, then the inflationary theory will in some way make its
presence clear in strong gravity astrophysical objects. Thus
studying neutron stars using inflationary models is of vital
importance because it may help in two things: firstly revealing
the correct modified gravity behind the observational reality and
secondly, inflationary models may be discriminated in neutron
stars. Also we have it in good authority that by studying multiple
inflationary models in neutron stars, we may pin down the optimal
equation of state for the neutron stars. We aim to report on this
issue very soon.

\section{Neutron Stars with Induced and Quadratic Inflationary Attractors}

We shall now use an astrophysics context and notation
\cite{Pani:2014jra} and we shall derive the
Tolman-Oppenheimer-Volkoff equations for the induced and quadratic
inflationary attractors. The physical units system we shall adopt
is the Geometrized units in which $G=c=1$. Let us translate the
expression for the gravitational action in the Jordan frame
developed in the previous section in an astrophysics context, so
we have,
\begin{equation}\label{ta}
\mathcal{S}=\int
d^4x\frac{\sqrt{-g}}{16\pi}\Big{[}\Omega(\phi)R-\frac{1}{2}g^{\mu
\nu}\partial_{\mu}\phi\partial_{\nu}\phi-U(\phi)\Big{]}+S_m(\psi_m,g_{\mu
\nu})\, ,
\end{equation}
and by performing a conformal transformation,
\begin{equation}\label{ta1higgs}
\tilde{g}_{\mu \nu}=A^{-2}g_{\mu \nu}\,
,\,\,\,A(\phi)=\Omega^{-1/2}(\phi)\, ,
\end{equation}
we get the Einstein frame action,
\begin{equation}\label{ta5higgs}
\mathcal{S}=\int
d^4x\sqrt{-\tilde{g}}\Big{(}\frac{\tilde{R}}{16\pi}-\frac{1}{2}
\tilde{g}_{\mu \nu}\partial^{\mu}\varphi
\partial^{\nu}\varphi-\frac{V(\varphi)}{16\pi}\Big{)}+S_m(\psi_m,A^2(\varphi)g_{\mu
\nu})\, ,
\end{equation}
with $\varphi$ being the Einstein frame canonical scalar field and
\begin{equation}\label{potentialns1}
V(\varphi)=\frac{U(\phi)}{\Omega^2}\, .
\end{equation}
The important quantities for the Einstein frame extraction of the
Tolman-Oppenheimer-Volkoff equations are the potential in the
Einstein frame, and the function $A(\varphi)=\Omega^{-1/2}(\phi)$.
Let us quote the expressions of $A(\varphi)$ expressions for the
induced and quadratic inflationary attractors, and in the case of
the induced inflation case we have,
\begin{equation}\label{inducedA}
 A(\varphi)=e^{-\frac{1}{2}\sqrt{\frac{2}{3}}\varphi}\, ,
\end{equation}
and also for this case, the function $\alpha (\phi)$ defined as,
\begin{equation}\label{alphaofvarphigeneraldef}
\alpha(\varphi)=\frac{d \ln A(\varphi)}{d \varphi}\, ,
\end{equation}
is equal to,
\begin{equation}\label{alphaofphifinalintermsofvarphi}
a(\varphi)=-\frac{1}{2}\sqrt{\frac{2}{3}}\, .
\end{equation}
Finally the scalar potential for the Einstein frame induced
inflation theory is,
\begin{equation}\label{indastreinstein}
V(\varphi)=V_s\left(1-e^{-\sqrt{\frac{2}{3}}\varphi} \right)^2\, .
\end{equation}
Now let us turn our focus on the quadratic attractors case in the
Einstein frame. In this case, the scalar potential in the Einstein
frame reads,
\begin{equation}\label{indastreinstein}
V(\varphi)=\lambda g_1^2\varphi^2\, ,
\end{equation}
while the function $A(\varphi)$ reads,
\begin{equation}\label{inducedA}
 A(\varphi)=\left(1+\xi g_1\varphi \right)^{-1/2}\, ,
\end{equation}
and finally the function $\alpha (\varphi)$ defined in Eq.
(\ref{alphaofphifinalintermsofvarphi}) in the quadratic attractors
case, is equal to,
\begin{equation}\label{alphaofphifinalintermsofvarphi}
a(\varphi)=-\frac{1}{2}\frac{\xi g_1}{1+\xi g_1 \varphi}\, .
\end{equation}
\begin{figure}[h!]
\centering
\includegraphics[width=30pc]{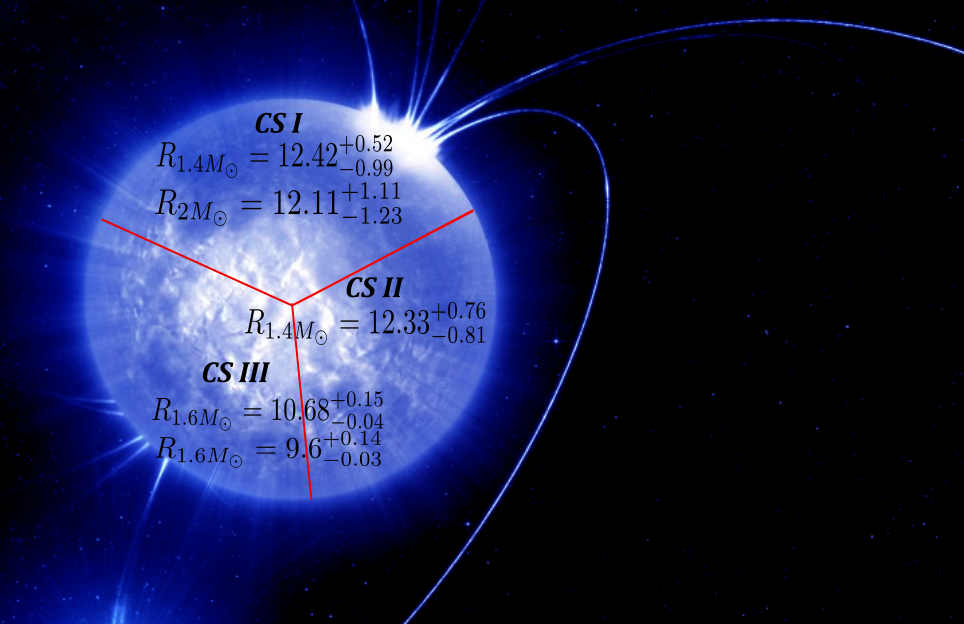}
\caption{The constraints CSI \cite{Altiparmak:2022bke}, CSII
\cite{Raaijmakers:2021uju}
 and CSIII \cite{Bauswein:2017vtn}. This figure is inspired from Credit: ESO/L.Cal\c{c}ada: \url{https://www.eso.org/public/images/eso0831a/.}} \label{plotcs}
\end{figure}
For the numerical analysis, in the case of induced inflation we
shall use the values $\xi\sim \mathcal{O}(10^{5})$, $\lambda\sim
\mathcal{O}(1)$ and for the quadratic inflationary attractor we
shall use $\xi\sim \mathcal{O}(10^{-2})$, $\lambda\sim
\mathcal{O}(1)$ and $g_1\sim \mathcal{O}(10^{-4})$, in order to
have a viable phenomenology. Contrary to what is usually done in
similar astrophysical works in the literature, the free parameters
of specific non-minimally coupled theories in the Einstein frame,
including the $R^2$ model, are not free parameters of the theory.
As we showed earlier these parameters are constrained by the
inflationary theory, so these must take specific values.
\begin{table}[h!]
  \begin{center}
    \caption{\emph{\textbf{CSI vs the Induced and Quadratic Attractors for the SLy, APR and WFF1 Equations of State for neutron stars Masses $M\sim 2M_{\odot}$}}}
    \label{table1}
    \begin{tabular}{|r|r|r|r|}
     \hline
      \textbf{Model}   & \textbf{APR} & \textbf{SLy} & \textbf{WFF1}
      \\  \hline
      \textbf{Quadratic Attractors Masses} & $M_{APR}= 0.709\,M_{\odot}$ & $M_{SLy}= 2.119\, M_{\odot}$ & $M_{WFF1}= 0.290\,
M_{\odot}$
\\  \hline
       \textbf{Quadratic Attractors Radii} & $R_{APR}= 11.414$km & $R_{SLy}= 10.898$km
      &$R_{WFF1}= 11.468$km
      \\  \hline
      \textbf{Induced Inflation Attractors Masses} & $M_{APR}= 2.018\, M_{\odot}$ & $M_{SLy}= 2.017\, M_{\odot}$ & $M_{WFF1}= 0.340\, M_{\odot}$
      \\  \hline
       \textbf{Induced Inflation Attractors Radii} & $R_{APR}= 11.155$km &
$R_{SLy}= 11.291$km
      & $R_{WFF1}= 11.069$km \\  \hline
    \end{tabular}
  \end{center}
\end{table}
The spacetime metric which describes static neutron stars is,
\begin{equation}\label{tov1}
ds^2=-e^{\nu(r)}dt^2+\frac{dr^2}{1-\frac{2
m(r)}{r}}+r^2(d\theta^2+\sin^2\theta d\phi^2)\, ,
\end{equation}
with the function $m(r)$ describing the gravitational mass of the
neutron stars and $r$ is the circumferential radius. Our aim is to
find numerically the functions $\nu(r)$ and $\frac{1}{1-\frac{2
m(r)}{r}}$. The procedure is the following, the function $\nu(r)$
is assumed to have a non-zero central value at the center of the
star.
\begin{figure}[h!]
\centering
\includegraphics[width=23pc]{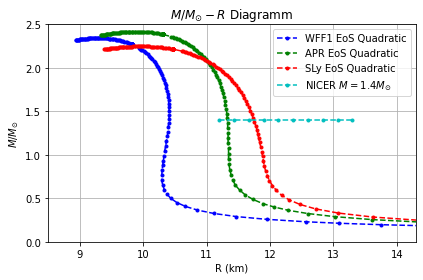}
\caption{The $M-R$ graphs for the quadratic attractor model for
the WFF1, APR and SLy equations of state.} \label{quadplot1}
\end{figure}
Beyond the surface of the star, in contrast to the General
Relativity case, there is no matching with the Schwarzschild
metric, because the metric functions receive contribution from the
scalar field effects, thus the only matching of the metric above
with the Schwarzschild metric will be done at numerical infinity.
Indeed, the presence of the scalar potential, and of the function
$A(\varphi)$ crucially affects the metric functions beyond the
surface of the star, thus no matching at the surface of the star
is required. The whole scalarization procedure is based on the
effects of the scalar field beyond the surface of the star.
\begin{table}[h!]
  \begin{center}
    \caption{\emph{\textbf{CSI vs the Induced and Quadratic Attractors for the SLy, APR and WFF1 Equations of State for neutron stars Masses $M\sim 1.4M_{\odot}$}}}
    \label{table2}
    \begin{tabular}{|r|r|r|r|}
     \hline
      \textbf{Model}   & \textbf{APR} & \textbf{SLy} & \textbf{WFF1}
      \\  \hline
      \textbf{Quadratic Attractors Masses} & $M_{APR}= 0.650\,M_{\odot}$ & $M_{SLy}= 1.425\, M_{\odot}$ & $M_{WFF1}= 0.258\,
M_{\odot}$
\\  \hline
       \textbf{Quadratic Attractors Radii} & $R_{APR}= 11.455$km & $R_{SLy}= 11.733$km
      &$R_{WFF1}= 11.951$km
      \\  \hline
      \textbf{Induced Inflation Attractors Masses} & $M_{APR}= 0.620\, M_{\odot}$ & $M_{SLy}= 1.392\, M_{\odot}$ & $M_{WFF1}= 0.270\, M_{\odot}$
      \\  \hline
       \textbf{Induced Inflation Attractors Radii} & $R_{APR}= 11.487$km &
$R_{SLy}= 11.777$km
      & $R_{WFF1}= 11.892$km \\  \hline
    \end{tabular}
  \end{center}
\end{table}
The Tolman-Oppenheimer-Volkoff equations for the scalar-tensor
theory in the Einstein frame are,
\begin{equation}\label{tov2}
\frac{d m}{dr}=4\pi r^2
A^4(\varphi)\varepsilon+\frac{r}{2}(r-2m(r))\omega^2+4\pi
r^2V(\varphi)\, ,
\end{equation}
\begin{equation}\label{tov3}
\frac{d\nu}{dr}=r\omega^2+\frac{2}{r(r-2m(r))}\Big{[}4\pi
A^4(\varphi)r^3P-4\pi V(\varphi)
r^3\Big{]}+\frac{2m(r)}{r(r-2m(r))}\, ,
\end{equation}
\begin{equation}\label{tov4}
\frac{d\omega}{dr}=\frac{4\pi r
A^4(\varphi)}{r-2m(r)}\Big{(}\alpha(\varphi)(\epsilon-3P)+
r\omega(\epsilon-P)\Big{)}-\frac{2\omega
(r-m(r))}{r(r-2m(r))}+\frac{8\pi \omega r^2 V(\varphi)+r\frac{d
V(\varphi)}{d \varphi}}{r-2 m(r)}\, ,
\end{equation}
\begin{equation}\label{tov5}
\frac{dP}{dr}=-(\epsilon+P)\Big{[}\frac{1}{2}\frac{d\nu}{dr}+\alpha
(\varphi)\omega\Big{]}\, ,
\end{equation}
\begin{equation}\label{tov5newfinal}
\omega=\frac{d \varphi}{dr}\, ,
\end{equation}
and recall that the function $\alpha (\varphi)$ is firstly defined
in Eq. (\ref{alphaofvarphigeneraldef}). With regard to the
pressure $P$ and the energy density $\epsilon$, we need to note
that these are Jordan frame quantities. Another notable feature is
that the spacetime in the interior and the exterior of the neutron
star is uniformly described by the spherically symmetric metric
(\ref{tov1}). However, the Tolman-Oppenheimer-Volkoff equations
shall be solved for the interior and the exterior of the star,
since beyond the surface, the pressure and energy density are
zero. The set of the initial conditions we shall use is the
following,
\begin{equation}\label{tov8}
P(0)=P_c\, ,\,\,\,m(0)=0\, , \,\,\,\nu(0)\, ,=-\nu_c\, ,
\,\,\,\varphi(0)=\varphi_c\, ,\,\,\, \omega (0)=0\, .
\end{equation}
We shall use a double shooting method in order to determine the
values of the metric function $\nu_c$ and $\varphi_c$ at the
center of the star, by optimizing the parameter values using as a
rule the optimal decay of the scalar field at the numerical
infinity. We shall use three distinct piecewise polytropic
equations of state \cite{Read:2008iy,Read:2009yp} with the low
density part corresponding to the SLy, WFF1 or the APR equations
of state. Since the temperature of neutron stars is much lower
than the Fermi energy of the particles that constitute neutron
stars, the neutron star matter can be described by a one-parameter
equation of state that accurately governs cold matter beyond the
nuclear density. The problem that arises though is that the
uncertainty in the equation of state is large, with the pressure
as function of the baryon mass density being uncertain to at least
one order of magnitude beyond the nuclear density. Furthermore,
the actual phase of matter in the core of neutron stars is also
uncertain. Thus a parameterized high-density equation of state can
be helpful and an optimal choice of an equation of state is a
high-density parameterized equation of state. This piecewise
polytropic equation of state, take into account astrophysical
phenomenological constraints regarding the nature of neutron stars
matter. The piecewise polytropic equation of state we shall use in
this paper are also take into account the causality constraints,
so causality is respected, see Refs. [111,112] for further
details. In general a piecewise polytropic equation of state can
be constructed by using a low-density part $\rho<\rho_0$, which is
usually chosen to be a well-known and tabulated crust equation of
state, and also the piecewise polytropic equation of state has a
high density part $\rho\gg \rho_0$. The numerical analysis we
shall perform shall yield the Einstein frame mass of the neutron
star, and from this we shall calculate the ADM mass in the Jordan
frame, using the following formula
\cite{Odintsov:2021qbq,Odintsov:2021nqa,Oikonomou:2021iid},
\begin{equation}\label{jordanframeADMmassfinal}
M_J=A(\varphi(r_E))\left(M_E-\frac{r_E^{2}}{2}\alpha
(\varphi(r_E))\frac{d\varphi
}{dr}\left(2+\alpha(\varphi(r_E))r_E\frac{d \varphi}{dr}
\right)\left(1-\frac{2 M_E}{r_E} \right) \right)\, .
\end{equation}
with $r_E$ being the radius of the neutron star in the Einstein
frame asymptotically and in addition $\frac{d\varphi
}{dr}=\frac{d\varphi }{dr}\Big{|}_{r=r_E}$. Also, the Jordan and
Einstein circumferential radii of the neutron star are related as
follows,
\begin{equation}\label{radiussurface}
R=A(\varphi(R_s))\, R_s\, .
\end{equation}
For notational simplicity, we shall denote with $M$ the Jordan
frame mass of the star, measured in solar masses $M_{\odot}$ and R
denotes the Jordan frame radius expressed in kilometers.
\begin{figure}[h!]
\centering
\includegraphics[width=23pc]{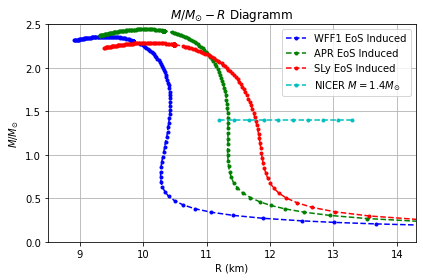}
\caption{The $M-R$ graphs for the induced inflation attractor
model for the WFF1, APR and SLy equations of state.}
\label{indplot1}
\end{figure}

\subsection{Results of the Numerical Analysis}

In this subsection we present the results of our numerical
analysis. Our aim is to find the $M-R$ graphs for the two
inflationary attractor models (Jordan frame quantities) and to
confront the data of the $M-R$ graphs with existing constraints on
neutron stars masses and radii which were developed nearly after
the GW170817 event. We will consider five in total constraints
which we classify in three distinct constraints, to which we will
refer to as CSI, CSII and CSIII. The CSI was developed in Ref.
\cite{Altiparmak:2022bke} and constrains the radius of an
$1.4M_{\odot}$ mass neutron star as
$R_{1.4M_{\odot}}=12.42^{+0.52}_{-0.99}$ and the radius of an
$2M_{\odot}$ mass neutron star as
$R_{2M_{\odot}}=12.11^{+1.11}_{-1.23}\,$km. The constraint CSII
was developed in Ref. \cite{Raaijmakers:2021uju} and constrains
the radius of an $1.4M_{\odot}$ mass neutron star to be
$R_{1.4M_{\odot}}=12.33^{+0.76}_{-0.81}\,\mathrm{km}$. Finally the
constraint CSIII is developed in Ref. \cite{Bauswein:2017vtn} and
constrains the radius of an $1.6M_{\odot}$ mass neutron star to be
larger than $R_{1.6M_{\odot}}=12.42^{+0.52}_{-0.99}\,$km while the
radius of a neutron star corresponding to the maximum mass must be
larger than $R_{M_{max}}>10.68^{+0.15}_{-0.04}\,$km. For reading
convenience in Fig. \ref{plotcs} we present the pictorial
representation of the constraints CSI, CSII and CSIII on neutron
stars.
\begin{table}[h!]
  \begin{center}
    \caption{\emph{\textbf{CSII vs the Induced and Quadratic Attractors for the SLy, APR and WFF1 Equations of State for Neutron Star Masses $M\sim 1.4M_{\odot}$}}}
    \label{table3}
    \begin{tabular}{|r|r|r|r|}
     \hline
      \textbf{Model}   & \textbf{APR} & \textbf{SLy} & \textbf{WFF1}
      \\  \hline
      \textbf{Quadratic Attractors Masses} & $M_{APR}= 0.593\,M_{\odot}$ & $M_{SLy}= 1.425\, M_{\odot}$ & $M_{WFF1}= 0.258\,
M_{\odot}$
\\  \hline
       \textbf{Quadratic Attractors Radii} & $R_{APR}= 11.523$km & $R_{SLy}= 11.733$km
      &$R_{WFF1}= 11.951$km
      \\  \hline
      \textbf{Induced Inflation Attractors Masses} & $M_{APR}= 0.563\, M_{\odot}$ & $M_{SLy}= 1.392\, M_{\odot}$ & $M_{WFF1}= 0.270\, M_{\odot}$
      \\  \hline
       \textbf{Induced Inflation Attractors Radii} & $R_{APR}= 11.569$km &
$R_{SLy}= 11.777$km
      & $R_{WFF1}= 11.892$km \\  \hline
    \end{tabular}
  \end{center}
\end{table}
For the numerical analysis we shall employ a python 3 numerical
code (variant of pyTOV-STT code \cite{niksterg}), using the LSODA
integrator and a double shooting method for determining the
optimal values of the metric function $\nu_c$ and of the scalar
field $\varphi_c$ at the center of the star, which make the scalar
field and metric function values vanish at numerical infinity. The
numerical infinity is taken to be $r\sim 67.943$ km.
\begin{figure}[h!]
\centering
\includegraphics[width=20pc]{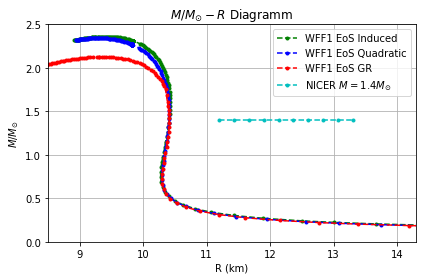}
\includegraphics[width=20pc]{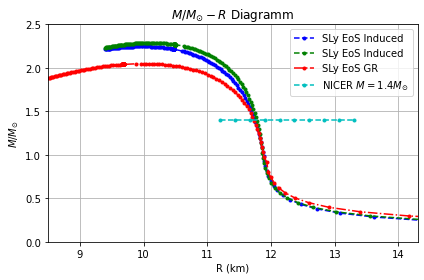}
\includegraphics[width=20pc]{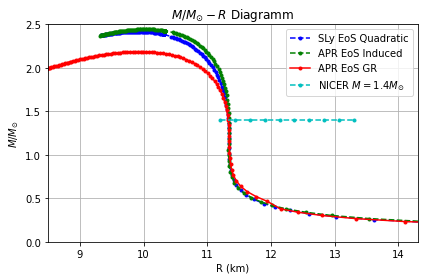}
\caption{The $M-R$ graphs for the quadratic, induced inflationary
attractor models and for the General Relativity case, for the
WFF1, APR and SLy equations of state.} \label{compplot}
\end{figure}
For all the plots that follow we added the NICER constraint for
$M=1.4M_{\odot}$ neutron stars which $90\%$ credible
\cite{Miller:2021qha} and constrains the neutron stars radius to
be $R=11.2-13.3\,$km. To start with, in Figs. \ref{quadplot1} and
\ref{indplot1} we present the $M-R$ graphs of the quadratic and
induced inflationary attractors neutron stars respectively, for
the WFF1 equation of state (blue curve), the APR equation of state
(green curve) and the SLy equation of state (red curve). As it is
apparent, the three distinct equations of state lead to different
maximum masses and radii, as it was expected. Also the largest
maximum mass is achieved for the stiffer APR equation of state.
Also a comparison of the models with the General Relativistic
result is given in the three plots of Fig. \ref{compplot} where we
present for each equation of state the $M-R$ graphs of the
quadratic attractors (blue curves), the induced attractors (green
curves) and the General Relativistic (red curves) for the WFF1
equation of state (upper left plot) the SLy equation of state
(upper right) and the APR equation of state (bottom plot). In all
cases the maximum masses are larger that those of General
Relativity, but the interesting part begins when the constraints
CSI, CSII and CSIII are considered.
\begin{table}[h!]
  \begin{center}
    \caption{\emph{\textbf{CSIII vs the Induced and Quadratic Attractors for the SLy, APR and WFF1 Equation of State for Neutron Star Masses $M\sim 1.6M_{\odot}$}}}
    \label{table4}
    \begin{tabular}{|r|r|r|r|}
     \hline
      \textbf{Model}   & \textbf{APR} & \textbf{SLy} & \textbf{WFF1}
      \\  \hline
      \textbf{Quadratic Attractors Masses} & $M_{APR}= 1.619\,M_{\odot}$ & $M_{SLy}= 1.584\, M_{\odot}$ & $M_{WFF1}= 1.606\,
M_{\odot}$
\\  \hline
       \textbf{Quadratic Attractors Radii} & $R_{APR}= 11.285$km & $R_{SLy}= 11.645$km
      &$R_{WFF1}= 11.115$km
      \\  \hline
      \textbf{Induced Inflation Attractors Masses} & $M_{APR}= 1.584\, M_{\odot}$ & $M_{SLy}= 1.609\, M_{\odot}$ & $M_{WFF1}= 1.075\, M_{\odot}$
      \\  \hline
       \textbf{Induced Inflation Attractors Radii} & $R_{APR}= 11.320$km &
$R_{SLy}= 11.680$km
      & $R_{WFF1}= 10.423$km \\  \hline
    \end{tabular}
  \end{center}
\end{table}
The results of the confrontation of the inflationary attractors
models with the observational constraints are presented in Tables
\ref{table1}-\ref{table5}. In Tables \ref{table1}, \ref{table2} we
confront the model with the constraint CSI. As it is apparent, the
SLy equation of state respects all the constraints for both
models, however the WFF1 case violates all the constraints. With
regard the APR equation of state, for the induced inflation case
respects the second CSI constraint (see Table \ref{table2}) and
also satisfies the CSIII constraint (see Table \ref{table3}) but
violates the rest of the constraints. For the quadratic inflation
case, the APR equation of state satisfies only CSIII and violates
all the rest. For the WFF1 equation of state, the models violate
all the constraints. Thus in conclusion, the only equation of
state which provides a viable neutron star phenomenology that
respects all the imposed constraints for the induced and quadratic
inflationary attractor models is the SLy equation of state. It is
apparent that in the post-GW170817 era, a viable phenomenological
neutron star model has to pass quite stringent tests in order to
be considered viable. Our result indicates that for the models
studied, intermediate stiffness equations of state can provide a
viable phenomenology, but this is a model dependent result. As we
demonstrated, for the quadratic and induced inflation attractors,
the only viable result is obtained for the SLy equation of state.
This is in contrast with previous work on attractors, which more
equation of state could provide a viable phenomenology. The result
depends strongly on the shape of the $M-R$ graphs. It seems that
the viability is guaranteed for curves strongly bend to the right
for radii between 11$\,$km-13$\,$km. In a future work we shall
provide a large sample of attractor models and we shall verify
this argument in an explicit way.
\begin{table}[h!]
  \begin{center}
    \caption{\emph{\textbf{CSIII vs the Induced and Quadratic Attractors for the SLy, APR and WFF1 Equations of State for Maximum Neutron Star Masses}}}
    \label{table5}
    \begin{tabular}{|r|r|r|r|}
     \hline
      \textbf{Model}   & \textbf{APR} & \textbf{SLy} & \textbf{WFF1}
      \\  \hline
      \textbf{Quadratic Attractors Maximum Masses} & $M_{APR}= 2.417\,M_{\odot}$ & $M_{SLy}= 2.248\, M_{\odot}$ & $M_{WFF1}= 2.342\,
M_{\odot}$
\\  \hline
       \textbf{Quadratic Attractors Radii} & $R_{APR}= 9.899$km & $R_{SLy}= 9.967$km
      &$R_{WFF1}= 9.281$km
      \\  \hline
      \textbf{Induced Inflation Attractors Maximum Masses} & $M_{APR}= 2.018\, M_{\odot}$ & $M_{SLy}= 2.286\, M_{\odot}$ & $M_{WFF1}= 2.360\, M_{\odot}$
      \\  \hline
       \textbf{Induced Inflation Attractors Radii} & $R_{APR}= 11.155$km &
$R_{SLy}= 9.981$km
      & $R_{WFF1}= 9.428$km \\  \hline
    \end{tabular}
  \end{center}
\end{table}
Before closing this section we need to discuss the necessity of
finding the predictions of inflationary potentials on the tidal
deformability of neutron stars and also study the radial
perturbation effects and the stability of neutron stars in
general, by also taking into account the constraints of the
GW170817 event. This is a non-trivial task and could be the
subject of a distinct article for the various distinct
inflationary attractors. For some recent relevant work on the
stability properties and perturbations of neutron stars in
scalar-tensor and unimodular gravity, see Refs.
\cite{Brown:2022kbw} and \cite{Yang:2022ees} respectively.

\section*{Concluding Remarks}

In this work we investigated the neutron star phenomenology of two
inflationary attractor potentials, that of quadratic and induced
inflation attractor potentials. The two models are known of
providing distinct inflationary phenomenology, with the quadratic
attractors providing a non-viable inflationary phenomenology,
while the induced inflationary attractors provide a viable
inflationary phenomenology, identical to the $R^2$ model of
inflation in the Einstein frame. We extracted the
Tolman-Oppenheimer-Volkoff equations in the Einstein frame for
both models, and by using a Python 3 LSODA double shooting method
we evaluated the masses and radii of neutron stars. From the
results we obtained the Jordan frame masses and radii of the
neutron stars and we constructed the $M-R$ graphs for three
distinct equations of state, the WFF1, the SLy and the APR
equation of state. Accordingly, we confronted the models with
three constraints that constrain the radii of specific mass
neutron stars. Specifically, we considered the three following
constraints, the CSI, CSII and CSIII. The CSI is studied in Ref.
\cite{Altiparmak:2022bke} and constrains the radius of an
$1.4M_{\odot}$ mass neutron stars to be
$R_{1.4M_{\odot}}=12.42^{+0.52}_{-0.99}$ and the radius of an
$2M_{\odot}$ mass neutron star to be
$R_{2M_{\odot}}=12.11^{+1.11}_{-1.23}\,$km. The constraint CSII
was studied in Ref. \cite{Raaijmakers:2021uju} and constrains the
radius of an $1.4M_{\odot}$ mass neutron star to be
$R_{1.4M_{\odot}}=12.33^{+0.76}_{-0.81}\,\mathrm{km}$. Finally the
constraint CSIII was studied in Ref. \cite{Bauswein:2017vtn} and
constrains the radius of an $1.6M_{\odot}$ mass neutron star to be
larger than $R_{1.6M_{\odot}}=12.42^{+0.52}_{-0.99}\,$km while the
radius of a neutron star corresponding to the maximum mass must be
larger than $R_{M_{max}}>10.68^{+0.15}_{-0.04}\,$km. As we showed,
the SLy equation of state respects all the constraints for both
models, however the WFF1 case violates all the constraints. With
regard the APR equation of state, for the induced inflation case
respects the second CSI constraint and also satisfies the CSIII
constraint, but violates the rest of the constraints. For the
quadratic inflation case, the APR equation of state satisfies only
CSIII and violates all the rest. For the WFF1 equation of state,
the models violate all the constraints. Thus only the SLy equation
of state respects all the constraints for both the induced and
quadratic inflation models. Also let us further note that if the
two inflationary models are compared for each equation of state,
the resulting neutron star phenomenology is quite similar if not
almost identical, see for example Fig. \ref{compplot}. This is
contrast with the inflationary phenomenology of the two attractor
models, which is quite different, since the induced attractors
provide a viable inflationary phenomenology while the quadratic
attractors are not viable cosmologically. This behavior is not a
general rule though, since the opposite can occur, that is, two
models might be identical in their inflationary phenomenology,
while producing quite different neutron star phenomenology. Work
is in progress toward this research line. In general let us
comment that it seems that cosmologically indistinguishable models
might be discriminated using their neutron star phenomenology and
vice versa. Intriguingly enough, scalar models which are
indistinguishable at frequencies of primordial cosmological
perturbations modes near the Cosmic Microwave Background ones, can
be distinguishable in future gravitational waves experiments that
can probe a stochastic tensor background. This is the opposite in
spirit to what we demonstrated in this paper, the fact that
phenomenologically distinguishable theories, can be distinguished
using their neutron star predictions. We shall support this
argument in future works. We need to note that the present
framework is not advantageous over some other modified gravity, it
is one possible description of nature, among other modified
gravities in general.

With regard to this last perspective, an important comment is in
order. In the present work the choice of the function $f(\phi)$ in
Eq. (\ref{nnminimalandpotential}) is not specified and the
potential is actually independent of it and rather universal. This
is the major difference with the $\alpha$-attractors case, studied
for example in our previous work \cite{Odintsov:2021qbq}. Also, we
need to note that for the quadratic attractors case, these are
entirely different from the $\alpha$-attractors case, for an
inflationary phenomenology point of view since the
tensor-to-scalar ratio and the spectral index in Eq.
(\ref{spectralindexlargelalpha}) are different from the ones
corresponding to the $\alpha$-attractors. Another useful comment
to add is that, although the induced inflation and
$\alpha$-attractors have the exact same inflationary phenomenology
described by Eq. (\ref{spectralindexsmallalpha}), in spite the
fact that these are basically different inflationary theories with
the same phenomenology (this justifies the terminology
attractors), in neutron stars they lead to different $M-R$ graphs.
In a future work we shall directly point out this issue, namely,
the fact that although distinct inflationary theories are
indistinguishable at the inflationary phenomenology level, they
can be actually distinguished in neutron stars.

\section*{Acknowledgments}

This research has been is funded by the Committee of Science of
the Ministry of Education and Science of the Republic of
Kazakhstan (Grant No. AP14869238)

\end{document}